\documentclass[a4paper,12pt]{article}
\usepackage[cp1251]{inputenc}
\usepackage{graphics}
\usepackage{amssymb,amsmath}
\usepackage[english]{babel}
\usepackage{indentfirst}
\begin{document}
\bigskip

\centerline {\bf On cyclic activity of the Sun and solar-type stars}
\bigskip

\centerline {E. A. Bruevich $^{a}$ and G. S. Ivanov-Kholodnyj
$^{b}$}

\bigskip

\centerline {\it $^a$ Sternberg Astronomical Institute, Moscow State
 University,}
\centerline {\it Universitetsky pr., 13, Moscow 119992, Russia}\
\centerline  {\it $^b$ Pushkov Institute of Terrestrial Magnetism,}
\centerline  {\it Ionosphere and Radio Wave Propagation of Russian}
\centerline  {\it Academy of Science, Troitsk, Moscow Region,
Russia}\

\centerline {\it e-mail: $^a${red-field@yandex.ru}, $^b$
{gor@izmiran.trotsk.ru}}\

\bigskip
{\bf Abstract.}
 The cyclicity of 33
solar-type stars that are similar to 11-year and to the
quasi-biennial variations of solar radiation have studied. Our
calculation were based on new simultaneous observations of the flux
variations of the photospheric and chromospheric emissions of 33
solar-type stars and the Sun during the HK project that were
conducted over the last 20 years. The method of Fast Fourier
Transform  was applied to these observed data. In addition to the
known cyclic chromospheric emission variations of stars at the
11-year time scale, which were discovered at the Mount Wilson
Observatory, we found a recurrences on the quasi-biennial time
scale. The results of calculations of periods of the star's fluxes
variations at the 11-year and quasi-biennial time scales are
presented.


\vskip12pt
\centerline
{\bf1. Introduction}
\vskip12pt The
photospheric observations of stars with active atmospheres that have
been conducted regularly in the optical range since the middle of
the 20th century found out low-amplitude variability in some of
them, which is mainly caused by the presence of dark spots on the
surfaces of stars, such as sunspots. Later, at the Mount Wilson
Observatory an extensive program of simultaneous observations of
more than a hundred stars with solar-type chromospheric activity was
launched. The chromospheric activity indices for solar-type stars
were studied by ((Baliunas,
 Donahue \& Soon W. 1995; Lockwood {\it et al.} 2007) at Mount
Wilson observational program during 45 years, from 1965 to present
time.

In the framework of the HK project the relationship of the flux in
the centers of the H and K Ca II emission lines (396.8 and 393.4 nm
respectively) to the fluxes in the nearby continuum (400.1 and 390.1
nm respectively) is determined. This relationship  is called the
$S_{Ca II}$ index by authors of the project. It's known that the
$S_{Ca II}$ index is a good indicator of the chromospheric activity
of the Sun and stars.

Here, we use new data of simultaneous observations of photospheric
and chromospheric fluxes variations of the Sun and 33 solar-type
stars, obtained during the last 20 years at the Lowell (photometric
observations) and Smithsonian observatories (observations of the H
and K Ca II emission lines) of Stanford University during the HK
project We use these data for the calculation of the cyclic activity
periods of the  fluxes variations of the Sun and stars.

 Observations of stars give us information only about the
magnitude of radiation fluxes from the full disk. By choosing stars
that are closest in their characteristics to the Sun, we transfer
knowledge about the physical processes in the solar atmosphere to
these stars, which makes it possible to successfully interpret their
in radiation flux variations. Naturally, the Sun and every single
solar-like stars differ in their mass, average density, surface
temperature, the total area of sunspots, their contrast, the
dependence of the temperature of the photosphere and chromosphere on
the different altitudes, sizes and contrast of faculae, and many
other parameters.

Analysis of the observations of radiation flux variations of the Sun
and solar-type stars suggests that the structures of the atmospheres
of most of these stars are similar to that of the Sun. The
parameters that differ depend on the mass of a star, its spectral
class, age, chemical composition, and so on. Such an approach makes
it possible to successfully interpret the observations of variations
of radiation fluxes of many solar-type stars, considering the stars
as analogs of the Sun, particulary at different stages of evolution.

In this work we confirm the existence of cyclic variations of the
radiation in the quasi-biennial time scale for the majority of the
studied solar-type stars, as well as to determine the periods of
these variations.

\vskip12pt

\centerline {\bf2. Long-term changes in the activity of solar-type
stars}
 \vskip12pt
  In this paper we consider the
cyclic variation of Sun's and star's atmospheres fluxes at 11-year
and quasi-biennial time scales. Note than the duration of the
so-called 11-year solar activity varies from 8 to 15 years according
to many years of observations. This includes regular observations of
solar activity during the last one and half centuries, as well as
circumstantial evidence, including radionuclide and other types of
evidence on timescales of 1000 years. The main indices of solar
activity that characterize the study of the full solar disk
radiation are the Wolf numbers (the longest series of observations)
and the radio flux at 10,7 cm wavelength ($F_{10,7}$). Extra-earth
atmospheric observations in the ultraviolet and X-ray ranges of the
whole solar disk and of individual active regions (characterizing
the activity at different altitudes of the atmosphere) were started
in the 1960s. Unfortunately, these observations, which are important
for understanding the nature of the solar activity, are less regular
compared with ground-based observations of Wolf numbers and
$F_{10,7}$. They depend on the operation time of the corresponding
instruments in the circumterrestrial orbit. However, according to
numerous studies of solar activity indices, in particular, flows in
individual lines in short-wave region, all of them correlate well
with Wolf numbers and with a more objective index of activity
$F_{10,7}$.

For the HK project stars, periods of 11-year cyclic activity
(according to observations (Lockwood {\it et al.} 2007) during 40
years change a little
 for one stars and range from 7 to 20 years for various
stars with established activity (about 30\% of the total number of
stars studied in this project). By analogy with the detailed study
of formations on the Sun,
 the formations on the disk that are responsible for the atmospheric
 activity without flares of solar-type stars include spots and faculae in the photosphere, flocculi
 in the chromosphere, and prominences and coronal mass ejections in the corona.
 Areas where all these phenomena are observed together are known as centers of activity or active regions.
Thus, when analyzing radiation fluxes in the H and K Ca II lines
(which are sensitive to chromospheric activity) and flows from the
entire disc in broadband filters of the photospheric ubvy-system
close to standard UBV-system as was shown in (Lockwood {\it et al.}
2007) it's necessary to take into account for the Sun and the stars
the contribution from the existing disk's active regions. Active
regions are formed where strong magnetic fields emerge from
underneath the photosphere. In other words, the different
manifestations of the activities of the atmospheres of the Sun and
stars are the result of the magnetic fields evolution. This involves
both global and local magnetic fields. Their interaction with the
magnetized substance in subphotospheric layers of stars involved in
convective motion is also important. By their size and lifetime
active regions vary greatly. They can be observed from several hours
to several mouths.

It was shown that from all solar-type stars that belong to stars of
spectral classes F,G and K of the main sequence on the
Hertzsprung-Russell diagram, a regular cyclicity of chromospheric
activity of the solar type is more common for stars of the late
spectral classes G and K with sufficiently formed subphotospheric
convective zones as was shown in (Bruevich, Katsova \& Sokolov
2001). These stars rotate relatively slow on their axes (the
rotation period is about 25 - 45 days, in contrast to 3 -10 days for
stars with thin subphotospheric convective zones).

The Sun, a star of G2 class, rotates with the period of 25 days. The
rotation period $T_{rot}$ around their axes for a sample of studied
stars, their spectral classes, the quality of their 11-year
cyclicity and values of respective periods $T^{HK}_{11}$, according
to prior calculations made in (Baliunas,
 Donahue \& Soon W. 1995) are given
in Table. We mentioned that all the interpretations of variations in
radiations fluxes of stars are based on the assumption that
solar-type stars (and it is confirmed by all observations) have the
same active regions, which evolve in one or more periods of rotation
around their axes by the same laws that operate on the Sun.

In addition, the time dependence of variations of the radiation flux
from the Sun and stars, which takes into account the rotational
modulation (the time scale is about 1 month), on large scales (of
approximately several years) takes the form of a sine wave with a
period of $T_{11}$ corresponding to the 11-year scale of cyclic
activity. Based on the example of changes in radiation fluxes in the
H and K CaII lines for the Sun and stars (as we see in (Lockwood
{\it et al.} 2007) ) that the amplitude of the sine wave varies
slightly from cycle to cycle. The change in the amplitude of the
light curves of stars for each cycle depends on the contribution of
preexisting and already decayed active regions in the so-called
background radiation in chromospheric lines. An additional
contribution to the amplitude increased in chromospheric lines is
made by brightness increase from so-called chromospheric network at
the peak of a cycle.

\vskip12pt
\begin{figure}[h!]
 \centerline{\includegraphics{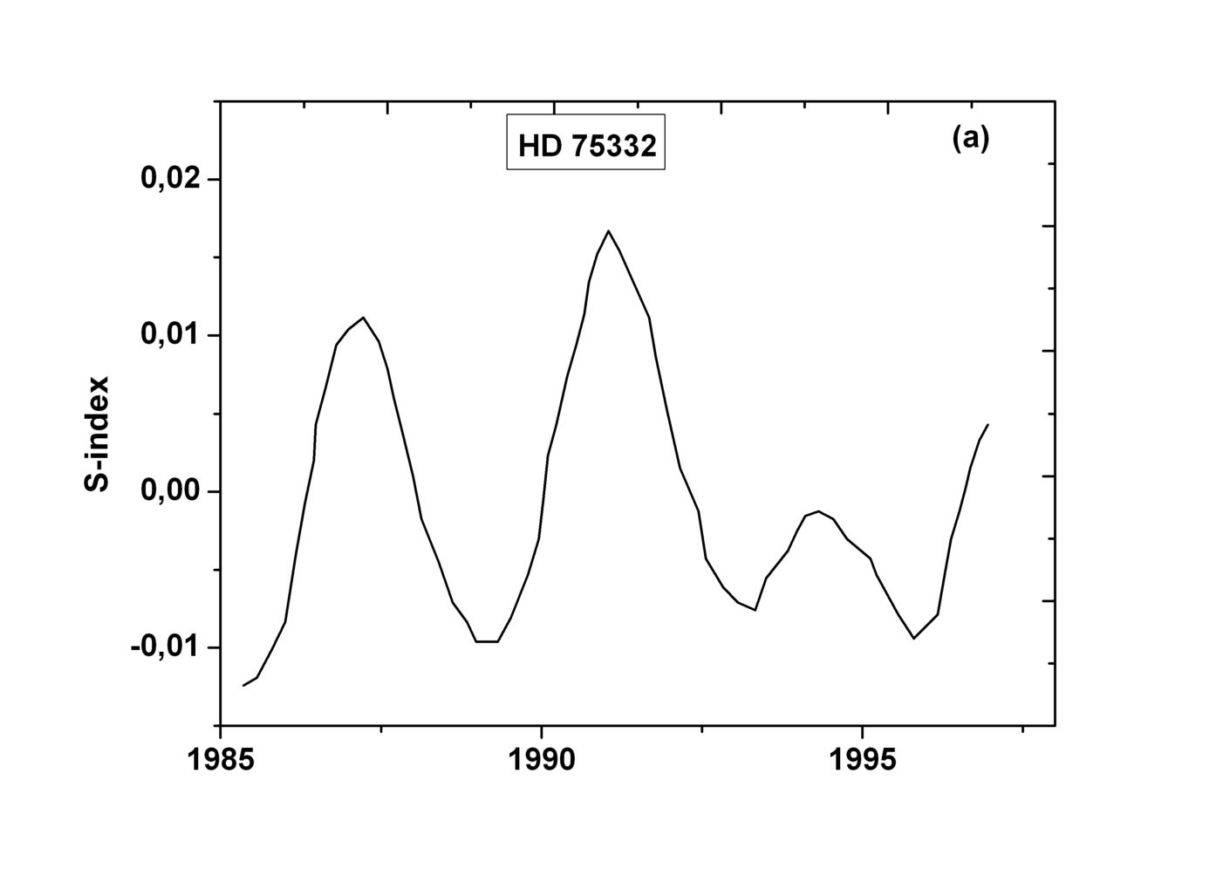}}
  \centerline{\includegraphics{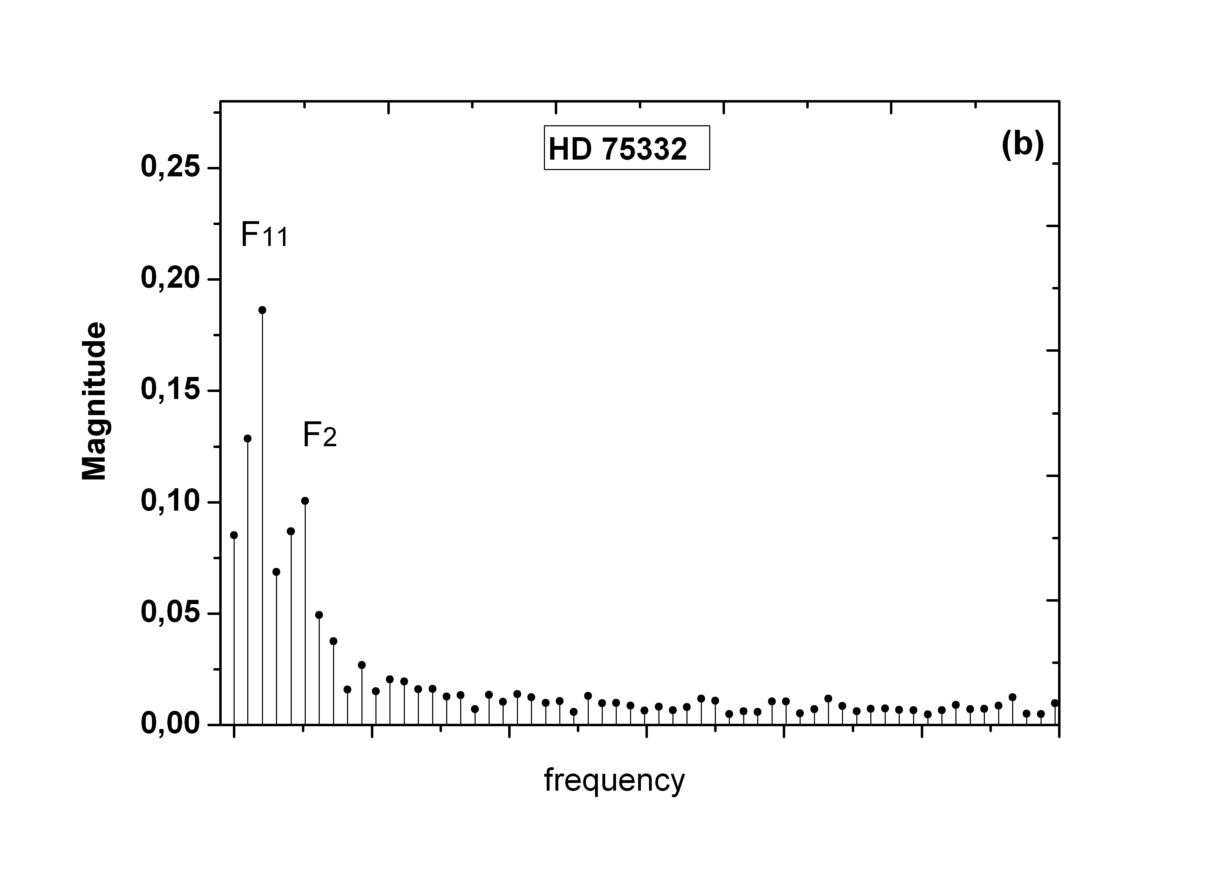}}
 \caption{Records of relative CaII emission fluxes (S-index) from the
Mount Wilson observations Lockwood {\it et al.} 2007 for  $HD
75332$ star.} \label{Fi:Fig1}

   \label{F-2panels}
\end{figure}
\vskip12pt

\vskip12pt
\begin{figure}[h!]
   \centerline{\includegraphics{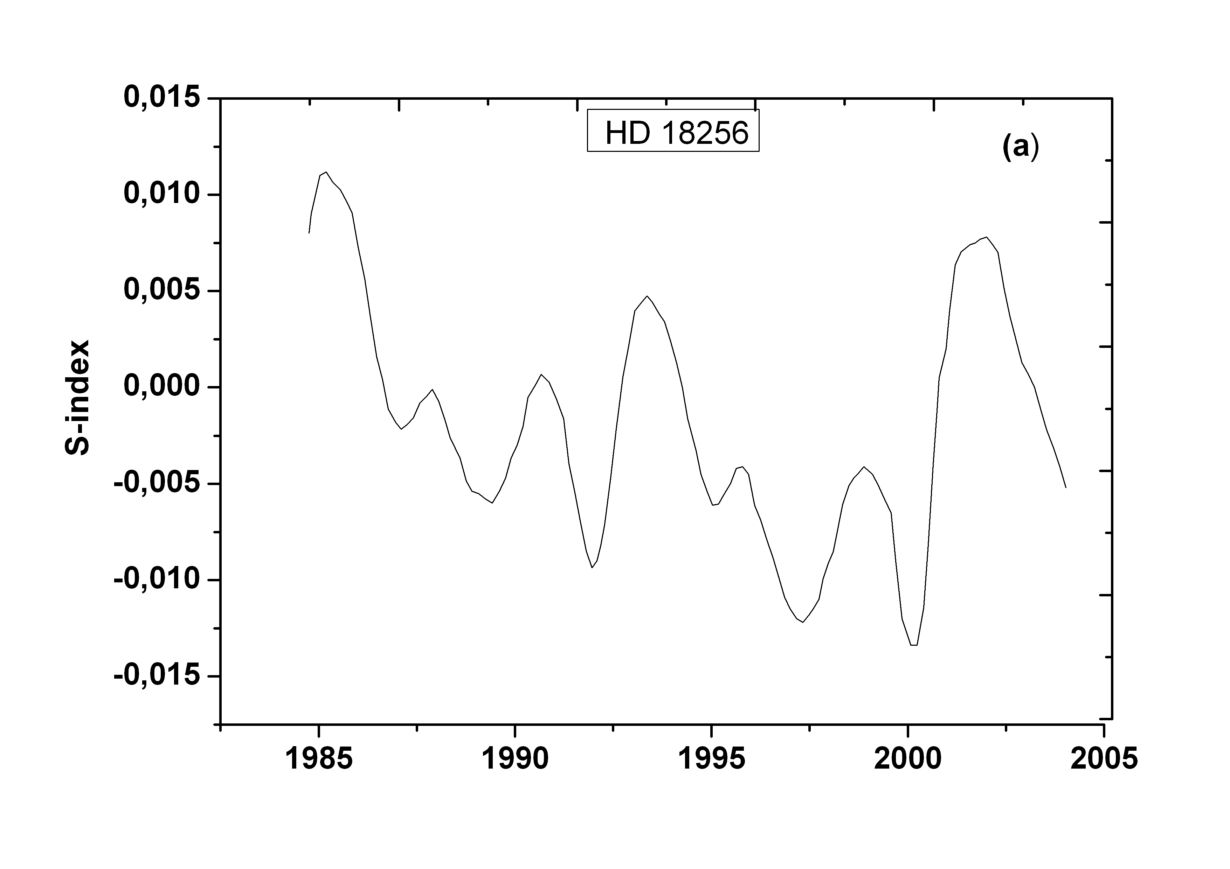}}
  \centerline{\includegraphics{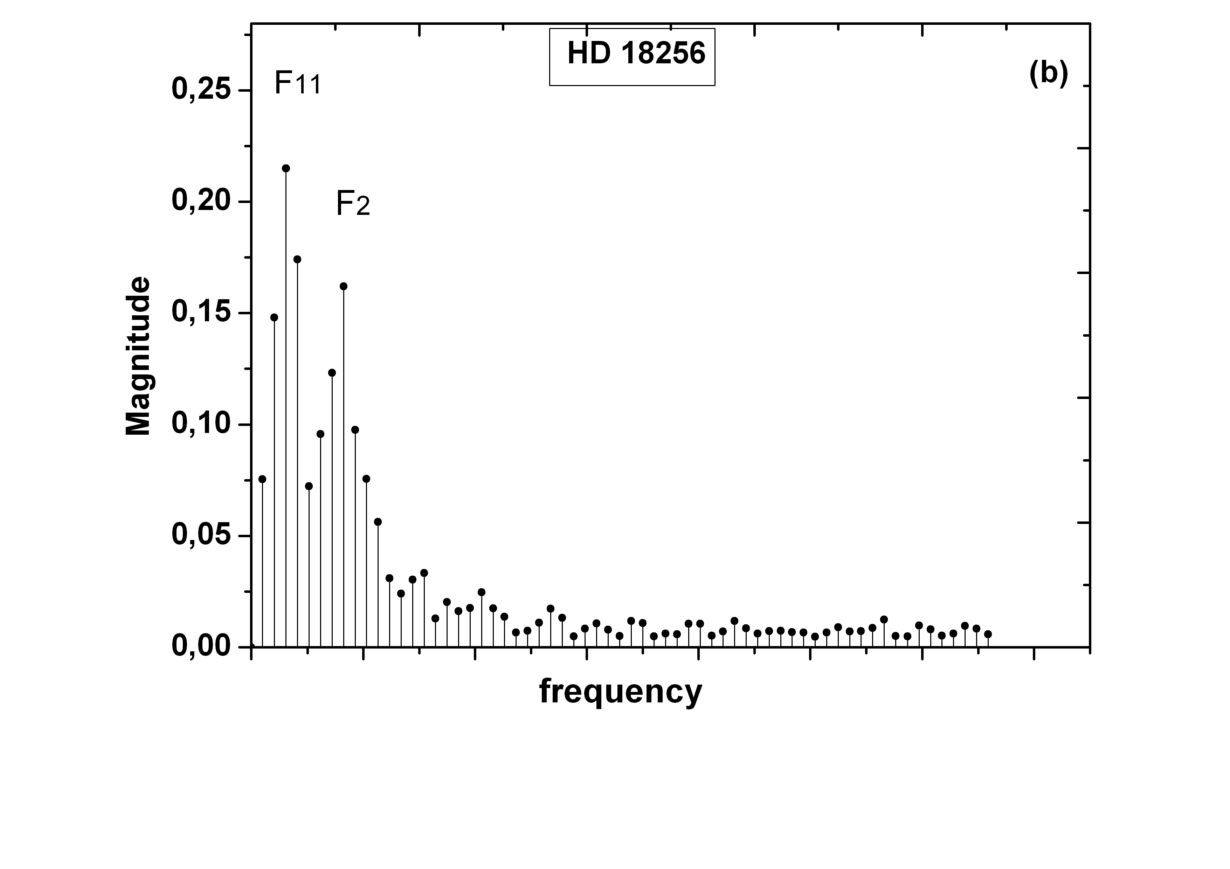}}
 \caption{Records of relative CaII emission fluxes (S-index) from the
Mount Wilson observations Lockwood {\it et al.} 2007 for  $HD
18256$ star.} \label{Fi:Fig2}

   \label{F-2panels}
\end{figure}

\centerline
{\bf3. Method and calculations}

\vskip12pt

 Solar and stellar activity is a combination of
regular manifestation of characteristic entities in the atmosphere,
which are associated with the release of large amounts of energy,
whose frequency and intensity change cyclically.

We use the Sun's and star's observed data taken from the (Lockwood
{\it et al.} 2007). In this work (for the first time) the results of
$S_{CaII}$ observations were published in the form of the smoothed
curves overlaid on the everyday observed data. These smoothed curves
were informative data which have been corrected with a glance of
Sun's and star's rotation.  (Kollath \& Olah 2009) used the wavelet
method for the determination of the EI Eri flux variations period on
quasi-biennial time scale. They used both the data which were
corrected with a glance of star's rotation (smoothed curves of
fluxes variations) and the data not corrected for star's rotation.
Their calculations showed that the use of the smoothed data curves
were preferred in such calculations because of smaller volume of
calculation actions without any losses of accuracy.

We used observations of radiation fluxes in chromospheric lines (the
$S_{CaII}$ index). We have scanned (with help of the Microlab Origin
7.5 program's package) the 33 star's and the Sun's observed data
from   (Lockwood {\it et al.} 2007). After the scanning of these
data ($S_{CaII}$ index observations of 33 stars and the Sun) we
obtained the observational series with equal time interval data.

Figures 1a and 2a show the scanned smoothed curve of the stars HD
75332 and HD  18256 (these stars we selected as illustration of our
calculations).

Then the method of Fast Fourier Transform (FFT) was applied to these
scanned data.

A FFT is an easy way to show the dominant frequencies in a signal.
We used the Microsoft Excel Data Analysis package to make a FFT
graphs.

In the Microsoft Excel we opened a blank spreadsheet.

We added the following titles to the first cells in columns A, B, C,
D and E of the spreadsheet: Time, Data, FFT Frequency, FFT Magnitude
and FFT Complex. We formed the columns of Time (first column A) and
Data (second column B - our scanned data).

In our case we took 128 points - our scanned data.  We used the
observed data obtained during about 20 years). The sampling rate by
the number of samples $V_N$ was equal to 128/19,6 points per year
for   HD 18256 calculations and equal to 128/20,2 for HD 75332.

In the "Time" column we determined the time at which each point was
taken. Than we performed a Fourier Analysis of our data with
commands: Data/Data Analysis/Fourier Analysis). The next columns C2
- C129 (FFT Frequency), D2 - D129 (FFT Magnitude) and E2 - E129 (FFT
Complex) were formed with help of these commands. The output data
formed the E column is called the FFT Complex because the Fourier
Analysis function outputs a complex number.

In the first cell of the "FFT Magnitude" column we typed the command
of the following equation:=IMABS(E2), then we dragged this equation
down so it fills every cell in this column. This action turned the
complex number in the "FFT Complex" column into a real number we can
use to find the dominant frequencies. Than we calibrated the X axis
of the graph to show the dominant frequencies.

We created a separate column (in our case column F) showing zero
through the number of data points minus one (N-1). We created a
separate cell with the following function:=(S/2)/(N/2) where S is
the sampling rate, and N is the Number of samples.

In the "FFT Frequency" column we typed the following
function=F2*$V_N$ where F2 is the appropriate number from the number
column, and $V_N$ denotes the function dividing the sampling rate by
the number of samples. Than we dragged this function down to half of
our data points (64 point only because of the overall number of data
points was equal to 128.)

Than we created a graph (see Figures 1b and 2b) of the "FFT
Magnitude" column (Y-axis) versus the "FFT Frequency" column
(X-axis). We used this method for the determination of  the periods
of 11-year cyclicity (the $T_{11}$) and of the quasi-biennial
cyclicity  (the $T_{2}$ period) of chromospheric activity for the
Sun and 33 stars. We also attempted to use the simultaneous
photometric observations scanned (Lockwood {\it et al.} 2007) data
of the 33 stars and the Sun for FFT calculations. But we understood
than $S_{CaII}$ index observed data (very sensitive to atmospheric
activity) have gone better to our purposes of determinations of
periods of fluxes variations.

At Figures 1b and 2b we see graphs of HD 75332 and HD  18256 FFT
calculation results with peaks where the dominant frequencies are
shown. The value $F_{11}$ corresponds to $T_{11}= 10,1$ and $T_{11}=
6,6$ years for HD 75332 and HD  18256, the value $F_{2}$ corresponds
to $T_{2}= 3,8$ and $T_{2}= 2,8$ year for HD 75332 and HD  18256
respectively.

\begin{figure}[h!]
 \centerline{\includegraphics{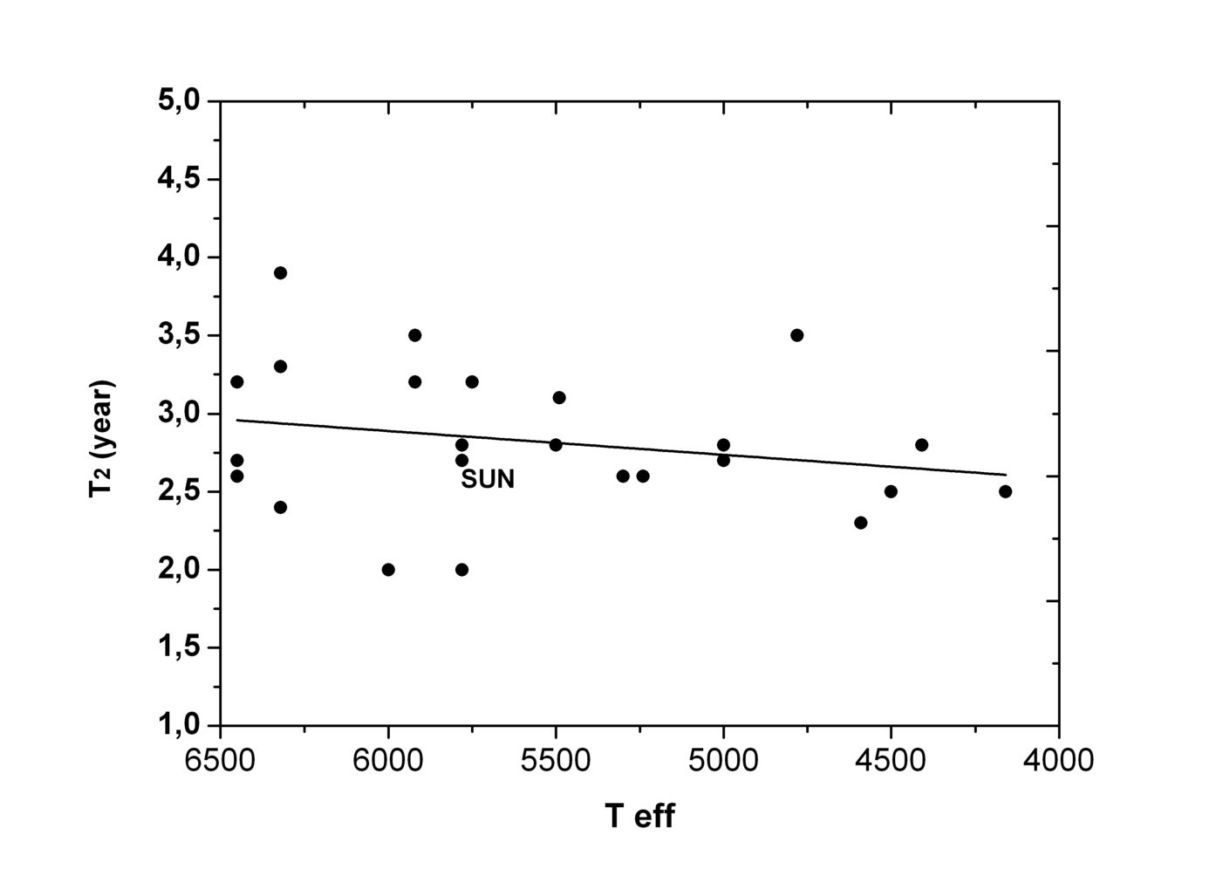}}

 \caption{ The dependence of $T_{2}$ versus $T_{eff}$}\label{Fi:Fig3}

\end{figure}

\begin{figure}[h!]
 \centerline{\includegraphics{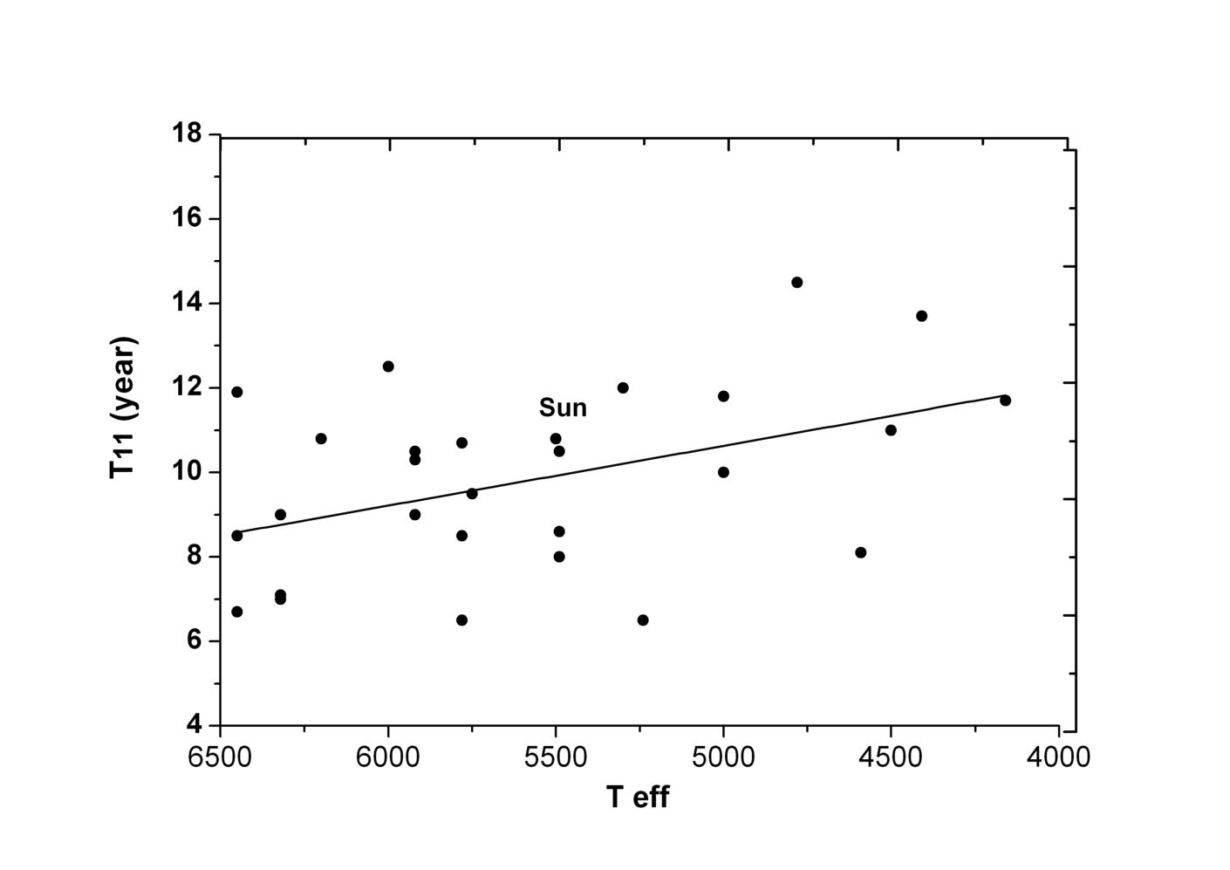}}

 \caption{ The dependence of $T_{11}$ versus $T_{eff}$}\label{Fi:Fig4}

\end{figure}

For studies of the Sun, one of the most important problems is to
examine the cyclical activity, which affects a number of terrestrial
processes. (Kvitova \&  Solanski 2008)  studied observed data of the
regular monitoring of solar irradiance (since 1978 t0 2008) and
showed that solar total and spectral irradiance varies at different
time scales. (Ivanov-Kholodnyj \& Chertoprud 2008) showed that the
importance of the QBV problem studying was emphasized in
solar-terrestrial processes. It turned out that QBV modulations of
total flux of solar radiation are closely related to various
quasi-biennial variations QBV processes on the Earth, in particular
with the QBV of the Earth's rotation speed, the QBV of the velocity
of the stratospheric wind, etc.

In our issue we looked for cyclic variations at different time
scales of HK project stars spectral irradiance. We also consider
that the stars of the HK project with stable cyclical activity in
the 11-year time scale, established after 40 years of continuous
observations, are so similar in their cyclical activity to the Sun,
that for the prediction of the radiation fluxes from these stars as
was showed in (Bruevich \& Bruevich 2004) methods that are used in
the practice of solar forecasting are quite applicable too. These
provisions confirm the similarity of the various manifestations in
the evolution of the atmospheric activity of the Sun and solar-type
stars. The existence of activity on the quasi-biennial time scale
for stars emphasized the fact that solar activity is a phenomenon
that is characteristic of the main sequence stars of late spectral
classes with developed subphotospheric convective zones. The correct
theory of the subphotospheric convection for Sun and solar-like
stars not having been made for the present time as was mentioned in
(Bruevich \& Rozgacheva 2010).

The results of our calculations are shown in the Table 1. We used
observations of radiation fluxes in chromospheric lines (the
$S_{CaII}$ index). Also we have to note than  $T_{11}$ and $T_{2}$
are the averaged values corresponding to the maximums of the
dominant frequencies at 11-year and quasi-biennial time scales.

A dash in columns with $T_{11}$ and $T_{2}$ means that we have not
identified the periodicity at a given time scale. The Table also
presents results of the determination of 11-year periods of stars
$T^{HK}_{11}$, determined for the first time by HK-project
participants and the quality of the cyclicity of the $S_{CaII}$
index variations from (Baliunas,
 Donahue \& Soon W. 1995).
One can see a good agreement of these results with our calculations
results of the $T_{11}$ periods.

Note that in some stars with well-determined 11-cyclicity (quality
of the cyclicity is EXCELL or GOOD, as mentioned in Table 1)
variations of fluxes are bimodal near the maximums of the cycle, in
a similar manner to the data for the Sun (according to Sun's and
star's observations from  (Lockwood {\it et al.} 2007)).

\begin{table}
\caption{Results of our calculations of $T_{11}$ and $T_{2}$ for 33
stars and the Sun}
\begin{tabular}{clclclclclclclcl} \\ \hline

  No & Star   &Spectral &$T_{rot}$ &Cyclicity       & $T_{eff}$&$T^{HK}_{11}$&$T_{11}$   &$T_{2}$ \\
    &         & class   & (days)   &class [1]&    &  (years)    & (years)   & (years) \\ \hline
   1 &  Sun   & G2-G4   &    25         &EXCELL        &     5780          & 10.0         &  10.7      &   2.7     \\
   2 &HD1835  & G 2.5   &     8         & FAIR         &     5750          & 9.1             &   9.5      &   3.2     \\
   3 &HD10476 & K1      &    35         &EXCELL        &     5000         &  9.6             & 10.0       &    2.8    \\
   4 &HD13421 & G0      &    17         &UNACT         &     5920        & -              & 10.0       &    -      \\
   5 &HD18256 & F6      &    3          &FAIR          &     6450         & 6.8             & 6.6        &   2.8     \\
   6 &HD25998  & F7     &     2         & UNACT        &     6320        &  -              &   7.1      &   -       \\
   7 &HD35296 & F8      &    4          &UNACT         &     6200         &  -              & 10.8        &   -       \\
   8 &HD39587 & G0      &    5          &UNACT         &     5920         &  -              & 10.0        &    -      \\
   9 &HD75332 & F7      &    4          & UNACT        &     6320         &  -               &   10.1         &   3.8    \\
   10 &HD76572 & F6      &   4          &POOR         &      6450         & 7.1              & 8.5        &    -     \\
   11 &HD81809  & G2     &   41         &EXCELL       &      5780         & 8.2              & 8.5        &   2.0      \\
   12 &HD82885 & G8      &   18         &FAIR         &      5490         & 7.9              & 8.6        &   -       \\
   13 &HD103095 & G8     &   31         &EXCELL       &      5490         & 7.3              & 8.0        &    -       \\
   14 &HD114710 & F9.5   &    12      &GOOD           &      6000         & 16.6             & 12.6         &   2.0     \\
   15 &HD115383 & G0     &   3          &UNACT        &      5920         &  -               & 10.3        &    3.5      \\
   16 &HD115104 & K1     &    18        &GOOD         &      5000         & 12.4             & 11.8         &   2.7     \\
   17 &HD120136 & F7    &   4          &POOR          &      6320         & 11.6             & -        &    3.3      \\
   18 &HD124570& F6      &    26        &UNACT        &      6450         & -                & -         &   2.7     \\
   19 &HD129333 & G0     &   3          &UNACT        &      5920        & -               & 9.0        &    3.2      \\
   20 &HD131156 & G2     &    6         &UNACT        &      5780         & -              & 8.5         &   2.8     \\
   21 &HD143761 & G0     &   17         &UNACT        &      5920         & -              &   -       &    -      \\
   22 &HD149661 & K2     &    21        &GOOD         &      4780         & 17.4           & 14.5      &   3.5     \\
   23 &HD152391 & G7   &     11         &EXCELL        &      5500        & 10.7           & 10.8        &    2.8      \\
   24 &HD157856 & F6     &    4        &FAIR          &       6450        & -             & 11.9        &   2.6     \\
   25 &HD158614 & G9     &   34         &UNACT       &       5300         & -              & 12.0      &    2.6      \\
   26 &HD160346 & K3     &    37        &EXCELL         &    4590         & 7.0             & 8.1         &   2.3     \\
   27 &HD161239 & G2     &   29         &FAIR        &       5780         & 5.7              & 6.5        &     -       \\
   28 &HD182572 & G8     &    41        &UNACT          &    5490         & -              & 10.5         &   3.1     \\
   29 &HD185144 & K0     &   27         &UNACT        &       5240        & -               & 6,5         &    2.6     \\
   30 &HD190007 & K4     &    29        &FAIR         &      4500         & 10.0             & 11.0         &   2.5     \\
   31 &HD201091 & K5     &   35         &EXCELL       &      4410         & 12.0             & 13.7        &    2.8      \\
   32 &HD201092 & K7     &    38        &GOOD         &      4160         & 12.4              & 11.7         &   2.5     \\
   33 &HD203387 & G8     &   -         &UNACT        &       5490         & -                &  -          &    2.6      \\
   34 &HD216385 & F7     &    7        &POOR         &       6320         & -               &  7.0         &   2.4     \\

\end{tabular}

\end{table}

\vskip12pt
\centerline {\bf4. Discussion}

\vskip12pt

The HK project  analyzed stars that belong to the solar type main
sequence stars of the F5 -K7 spectral classes. We see that the
values of quasi-biennial fluxes variations vary from 2,0 to 3,2
years. The observed data presented by (Lockwood {\it et al.} 2007)
showed that HK project stars have about 5 - 7 complete
quasi-biennial periods during the 20 years of observational time.

At Figure 3 we see dependence of quasi-biennial periods $T_2$ versus
spectral classes of stars (or  their $T_{eff}$).  We see that $T_2$
practically not depended from the spectral classes of stars.

Many authors are of the opinion that the value of solar fluxes
variations period in different spectral ranges (which is equal to 11
years at an average)  is determined by the size and the structure of
the subphotospheric convective zone. We assume that it's interesting
to analyze how the values of fluxes variations periods on the
11-year time scale differ for the stars studied here.

At Figure 4 we see  the  dependence of  11-year periods of  fluxes
variations  $T_{11}$  for the solar type main sequence stars of the
F5 -K7 spectral classes versus spectral classes of stars (or of
their $T_{eff}$). We see that the average values of these periods
increase upon to 30-40\% approximately for stars when their spectral
classes change from F5 to K7.

From the theory of internal structure of the stars it's known that
the value of the size of subphotospheric convective zone change from
$0,3 R^*$ (star's radius) for the Sun and the stars of late F-stars
and  G-stars to $0,5 R^*$ for  K  and M-stars. The results of our
calculations (see Figure 4.) confirm our  assumption that the
11-year periods of  fluxes variations  $T_{11}$  for the solar type
main sequence stars are connected with the sizes and structures of
these star's subphotospheric convective zones.

In the Table 1 we see that about 75\% of stars from the sample from
(Lockwood {\it et al.} 2007) have well-pronounced quasi-biennial
fluxes variations, similar to variation of solar radiation at the
same time scale. One can see that periods of quasi-biennial star's
cycles (values $T_{2}$) differ in the range from 2.2 to 3.8.

The preliminary analysis of the QBV of the star's chromospheric
fluxes showed that the duration of this quasi-biennial cycle is not
constant during one 11-year cycle, similar to the case of QBV of
Sun's fluxes. (Khramova, Kononovich \& Krasotkin 2002) showed that
the duration of the QBV of Sun's fluxes varies on average from 39
months at the beginning of the 11-year cycle to 25 months at the end
of it.

\vskip12pt
\vskip12pt
\centerline {\bf5. Conclusions}

\vskip12pt

 A statistical analysis was conducted of
observations of the Sun and 33 stars of the HK project using the
fast Fourier transform. From the spectral analysis of the
corresponding changes of radiation fluxes in Ca II lines we defined
the periods of the 11-year cyclicity  - $T_{11}$  and also revealed
a recurrence that similar to the quasi-biennial solar cyclicity and
determined the $T_{2}$ periods. The Table 1 shows that the periods
of cycles on the 11-year scale $T^{HK}_{11}$ identified by observers
at Mount Wilson during the primary treatment of the HK project
observations from (Baliunas,
 Donahue \& Soon W. 1995) agree well with our $T_{11}$ periods.

It can be concluded that the QBV phenomenon of radiation fluxes is
common among solar-type stars. According to our estimates, this QBV
phenomenon is observed in stars about twice as often as
well-determined cyclic activity in the 11-year scale. The periods of
QBV cycles $T_{2}$ for the studied sampling of stars, according to
our calculations, range from 2.2 to 3.8 years.

{\bf Acknowledgements} The authors thank the RFBR grant 09-02-01010
for support of the work.

\vskip12pt

\centerline {\bf References}

      Baliunas, S.L., Donahue R.A.  \& Soon W. 1995, {\it
Chromospheric variations in main-sequence stars, Astrophysical
        Journal}, {\bf 438}, 269.

    Bruevich, P.V. \& Bruevich E.A. 2004, {\it The possibility of simulation of the chromospheric
    variations for main-sequence stars, Astron. and Astrophys.
    Transactions}, {\bf 23}, N 2, 165, {\it ArXiv e-prints
    (arXiv:1012.5080v1)}.

     Bruevich, E.A., Katsova,  M.M. \& Sokolov D.D. 2001, {\it Levels of coronal
     and chromospheric activity in late-type stars and various types
      of dynamo waves, Astron. Report},  {\bf 78},
     827.

    Bruevich, E.A., \& Rozgacheva I.K. 2010, {\it On laminar convection
in solar tipe stars, ArXiv e-prints
 (arXiv:1012.3693v1)}.

    Ivanov-Kholodnyj, G.S. \& Chertoprud, V.E. 2008, {\it Quasi-biennial variations of the total solar flux:
  their manifestation in variations of the stratospheric wind and velocity of the earth rotation,
  Sol.-Zemn. Fiz.}, {\bf 2}, 291.

    Khramova, M.N., Kononovich, E.V. \& Krasotkin, S.A. 2002, {\it Quasi-biennial oscillations of global solar-activity
 indices, Astron. Vestn.}, {\bf 36},
548.

    Kollath, Z., Olah, K. 2009, {\it Multiple and changing cycles of active stars, ArXiv e-prints
    (arXiv:0904.1725v1)}.

    Kvitova N.A., Solanki S.K. 2008, {\it Models of solar irradiance
    variations: current status, J. Astrophys. Astr. {\bf 29}, 151.

     Lockwood, G.W., Skif, B.A., Radick R.R., Baliunas, S.L.,
 Donahue R.A. and Soon W. 2007,  {\it
  Astrophysical Journal Suppl.}, {\bf 171}, 260.

\end{document}